\documentclass[twocolumn]{aastex701}
\usepackage{amsmath,amssymb,amsfonts,array,graphicx}

\begin{document}

\title{Smoking-gun signatures of bounce cosmology from echoes of relic gravitational waves}

\author[orcid=0000-0001-9433-2871]{Mian Zhu}
\affiliation{College of Physics, Sichuan University, Chengdu 610065, China}
\email{zhumian@scu.edu.cn}

\author[orcid=0000-0003-0706-8465]{Yi-Fu Cai}
\affiliation{Department of Astronomy, School of Physical Sciences, University of Science and Technology of China, Hefei, Anhui 230026, China}
\affiliation{CAS Key Laboratory for Researches in Galaxies and Cosmology, School of Astronomy and Space Science, University of Science and Technology of China, Hefei, Anhui 230026, China}
\email{yifucai@ustc.edu.cn}
\correspondingauthor{Yi-Fu Cai}

\begin{abstract}
We report a novel feature of relic gravitational waves (GWs) in non-singular bounce cosmologies that is testable in light of GWs astronomy. In non-singular bounce cosmologies, the effective potential $M_p^2 a^{\prime \prime}/a$ that governs the evolution of primordial GWs contains two peaks due to the existence of contraction phase prior to the standard expansion phase. Accordingly, relic GWs interference between the two peaks in the effective potential. This interference results in a distinctive oscillatory feature in the energy density spectrum of GWs, analog to the resonant tunneling effect in quantum mechanics. As a result, the GWs spectrum exhibits an oscillatory patterns on high frequency regime, distinctive to other cosmological scenarios such as inflation. We show that the amplitude of GWs spectrum is high enough to reach the sensitivity of current and forthcoming GWs instruments, making our predictions falsifiable. Hence, our finding offers a promising way to experimentally test the non-singular bounce scenarios and search for new physics in early universe cosmologies.
\end{abstract}


\section{Introduction}

Non-singular bounce cosmologies \cite{Novello:2008ra, Battefeld:2014uga, Brandenberger:2016vhg} serve as a competitive early universe scenario alternative to inflation. By introducing a contracting phase prior to the standard Big Bang expansion, non-singular bounce cosmologies resolve the initial singularity \cite{Borde:1993xh, Borde:2001nh} and the trans-Planckian problem \cite{Martin:2000xs,Bedroya:2019snp} that puzzles inflation. Furthermore, bounce cosmologies can equally explain the formation of large-scale structure and the observed cosmic microwave backgrounds \cite{Finelli:2001sr, Cai:2012va, Cai:2014xxa, Cai:2014bea}, thus consistent with astrophysical observations \cite{Ageeva:2022asq,Ageeva:2024knc}.

Given the appealing feature of non-singular bounce cosmologies, it is natural to ask if it can be experimentally distinguished from inflation. In the phase of contraction, the energy density of anisotropic stress grows rapidly since it scales as $a^{-6}$ where $a$ is the scale factor of the universe. To evade the over-production of anisotropies, there must be a contracting phase with effective Equation-of-state (EoS) parameter $w_c > 1/3$ so that the background energy density $\rho_{\rm bg} \propto a^{-3(1+w_c)}$ grows faster than that of anisotropies \cite{Khoury:2001wf,Khoury:2001bz,Bozza:2009jx,Cai:2013vm,Cai:2013kja}. Primordial tensor fluctuations that cross the Hubble horizon in this phase has a strongly blue-tilted spectrum, i.e., the spectra index satisfying $2 < n_T < 3$. The existence of the blue spectrum at certain frequency regime is regarded as a crucial feature of bounce cosmologies, and have wide phenomenological implications, e.g., the interpretations of power deficit in the CMB TT spectrum \cite{Qiu:2015nha, Li:2016awk, Cai:2017pga}, the recent ACT observation \cite{Choi:2025qot}, and pulsar timing array (PTA) signals \cite{Li:2024oru, Li:2024dce, Qiu:2024sdd, Lai:2025xov, Lai:2025efh}. 
Nonetheless, inflation can as well yield a blue tensor spectrum \cite{Cai:2014uka, Cai:2020qpu, Choudhury:2023kam} at certain scales, see \cite{Wang:2014kqa} for a review. Hence, a blue-tilted primordial GWs spectrum can hardly be a deterministic feature of bounce scenarios. We are thus in a hot pursuit to search for distinctive features of bounce cosmologies. 

In this Letter, we report a unique pattern of primordial GWs in non-singular bounce cosmologies that can serve as a distinctive signature to experimentally test it in the GW astronomy. The production of primordial GWs can be understood as the scattering process of relic gravitons by the effective potential $M_p^2V(\tau) \equiv M_p^2 a^{\prime \prime}/a$. The effective potential has one peak in generic inflation models, and two peaks in non-singular bounce cosmologies due to the existence of contracting phase. This double-peak structure in non-singular bounce cosmologies results in the inteference of GWs between the two peaks in the effective potential, which is absent in inflation. Therefore, the GWs spectrum in non-singular bounce cosmologies has an  distinctive oscillatory pattern on high frequency regime that is not seen in generic inflation models. The amplitude of GWs spectrum on such scales is high enough to be detected by current and forthcoming GWs instruments. Our findings then provide a unique avenue to test bounce cosmologies and the underline new physics in GWs astronomy. 

\section{Primordial Gravitational Waves}

\begin{figure*}[ht]
    \centering
    \includegraphics[width=2.3in]{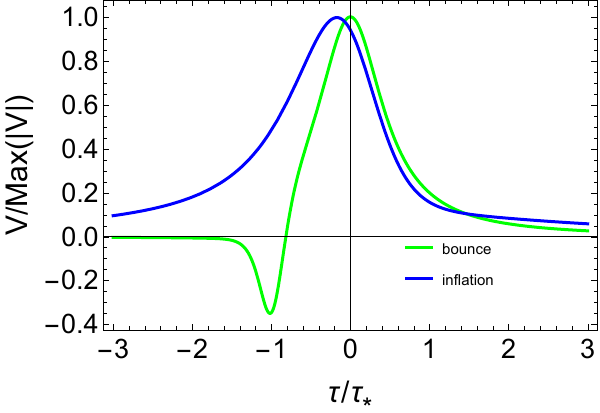}
    \includegraphics[width=2.3in]{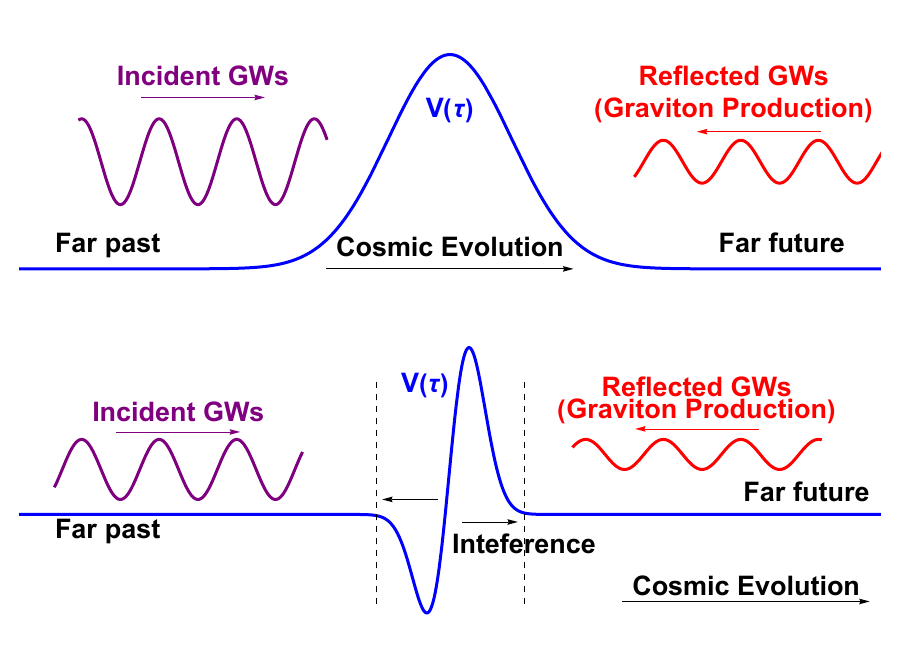}
    \includegraphics[width=2.3in]{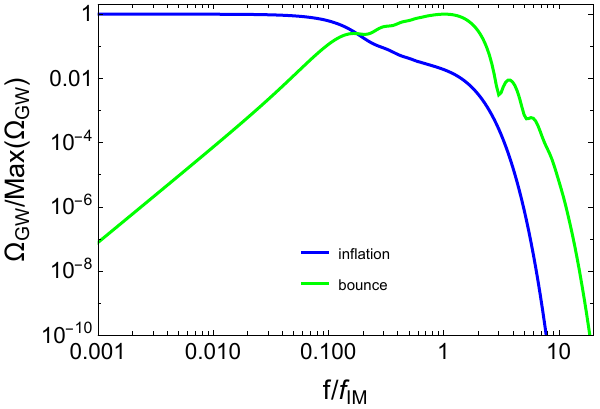}
    \caption{Left panel: the effective potential $M_p^2V(\tau)$ in inflationary (blue line) and bounce (green line) cosmologies. Middle panel: propogation of primordial GWs with different structure of effective potential (one-peak structure in the upper panel and two-peak structure in the lower panel). Right panel: the energy density spectrum in inflationary (blue line) and bounce (green line) cosmologies. Conformal time $\tau$ is rescaled such that $\tau = -\tau_{\ast}$($\tau = 0$) represents the location of negative(positive) peak in bounce scenario. $f_{\rm IM}$ is the frequency of GWs that cross the horizon at the end of inflationary epoch/contraction phase, respectively.}
    \label{fig:propo}
\end{figure*}

In a spatially flat Friedmann Lemaître Robertson Walker (FLRW) universe, primordial tensor fluctuations $h_k$ evolve as
\begin{align}
\label{eq:vkdynamical}
    v_k^{\prime \prime} + \left( k^2 - V(\tau) \right) v_k = 0 ~,
\end{align}
with $v_k \equiv a h_k/2$ the canonically normalized tensor fluctuations, $V(\tau) \equiv a^{\prime \prime}/a$, and a prime denotes differentiation with respect to conformal time $\tau$. In the far past, primordial fluctuations are generated on sub-horizon scales $k^2\gg V(\tau)$. The vacuum initial condition $v_k(\tau) = \frac{e^{-i k \tau}}{\sqrt{2k}}$ represents a plane wave propogate forward in time. In the present era, the GWs we can measure are only those with wavelengths shorter than the cosmological horizon $k^2\gg V(\tau)$. For those modes $v_k$ also evolves as a plane wave $v_k = \alpha_k \frac{e^{-i k \tau}}{\sqrt{2k}}+\beta_k \frac{e^{i k \tau}}{\sqrt{2k}}$. Thus, the evolution of primordial GWs can be understood as the scattering of a plane wave by a localized effective potential $M_p^2V(\tau)$, see Fig. \ref{fig:propo} for illustration.

Eq. \ref{eq:vkdynamical} is reminiscent to the famous one-dimentional potential well problem in quantum mechanics, where $V(\tau)$ plays the role of potential barrier. It is well-known that in quantum mechanics, resonant tunneling effect takes place when the potential barrier contains multiple peaks, which is absent in the case of single-peak potential barrier. The inteference of waves between the multiple peaks leads to oscillations in the reflected and transmitted waves. 

Likewise, the oscillatory pattern emergies in the spectrum of GWs when the effective potential develops multiple-peak structure. The double-peak structure of effective potential $V(\tau)$ acts as a temporal scattering cavity, and the primordial tensor mode is partially reflected by each peak in $V(\tau)$. Tensor modes are therefore partially reflected by both peaks, and the two reflected components acquire a relative $2k|(\tau_2-\tau_1)|$, where $\tau_1$ and $\tau_2$ denote the conformal-time positions of the two peaks, where $\tau_2$ and $\tau_1$ are the temporal location of the peaks in $V(\tau)$. Their superposition produces an interference term in $|\beta_k|^2$, leading to oscillatory features in $\Omega_{\rm GW}(f)$. Accordingly, the oscillation frequency in $\Omega_{\rm GW}(f)$ is $\Delta f \simeq [2a_0|\tau_2-\tau_1|]^{-1}$ where $a_0$ is the scale factor at today.

For illustrative purposes, we parametrize the double-peak structure as
\begin{equation}
\label{eq:VLorentz}
    V(\tau) = A_2^2 L(\tau; \tau_2,\Delta \tau) - A_1^2 L(\tau;\tau_1,\Delta \tau) ~,
\end{equation}
with $L(\tau; \tau',\Delta\tau)=\frac{1}{\pi}\frac{\Delta\tau}{(\tau-\tau')^2+\Delta\tau^2}$ the Lorentzian peak. On high frequency regime the Born approximation tells
\begin{align}
\label{eq:betakUV}
    \beta_k \simeq \frac{i e^{-2k \Delta \tau}}{2k} \left( A_2^2 e^{2ik\tau_2} - A_1^2 e^{2ik\tau_1} \right) ~,
\end{align}
so $|\beta_k|^2$ develops an oscillatory pattern in momentum space with period $\Delta k = \pi/|\tau_2 - \tau_1|$, or equivalently $\Delta f = [2a_0|\tau_2-\tau_1|]^{-1}$ in the observed GWs spectrum, as expected. We provide the technical details of \eqref{eq:betakUV} and related discussion in Appendix~\ref{app:oscillations}. 

In general, the effective potential contains one/two peak(s) in inflationary/bounce cosmologies, respectively (left panel of Fig. \ref{fig:propo}). As a result, on high-frequency regime, $|\beta_k|^2$ in bounce cosmologies obtain an additional oscillatory features. Notably, the observed energy density spectrum of relic gravitons is associated to $|\beta_k|^2$ via $\Omega_{\rm GW} = \frac{k^4 |\beta_k|^2}{3\pi^2M_p^2H^2a^4}$, with $\rho_c \equiv 3H_0^2 M_p^2$ the critial energy density and $H_0$ the Hubble parameter today. Therefore, bounce cosmologies predict an oscillatory pattern on high-frequency GWs that is absent in inflation (right panel of Fig. \ref{fig:propo}). This feature  can be distinguishable and serve as a smoking-gun signature of bounce cosmologies. 

Now we analyze the structure of effective potential in inflationary and bounce cosmologies. In inflationary epoch, $V(\tau) \simeq 2/\tau^2$, so the effective potential is initially positive and monotonously increases as the universe inflates from $\tau \to -\infty$ to $\tau \to 0$. After that, effective potential monotonously decreases in reheating phase and finally shrinks to 0 in the standard radiation domonated (RD) epoch. Thus, the effective potential generically contains only one peak in inflationary cosmologies.

\begin{figure}[h]
    \centering
    \includegraphics[width=0.95\linewidth]{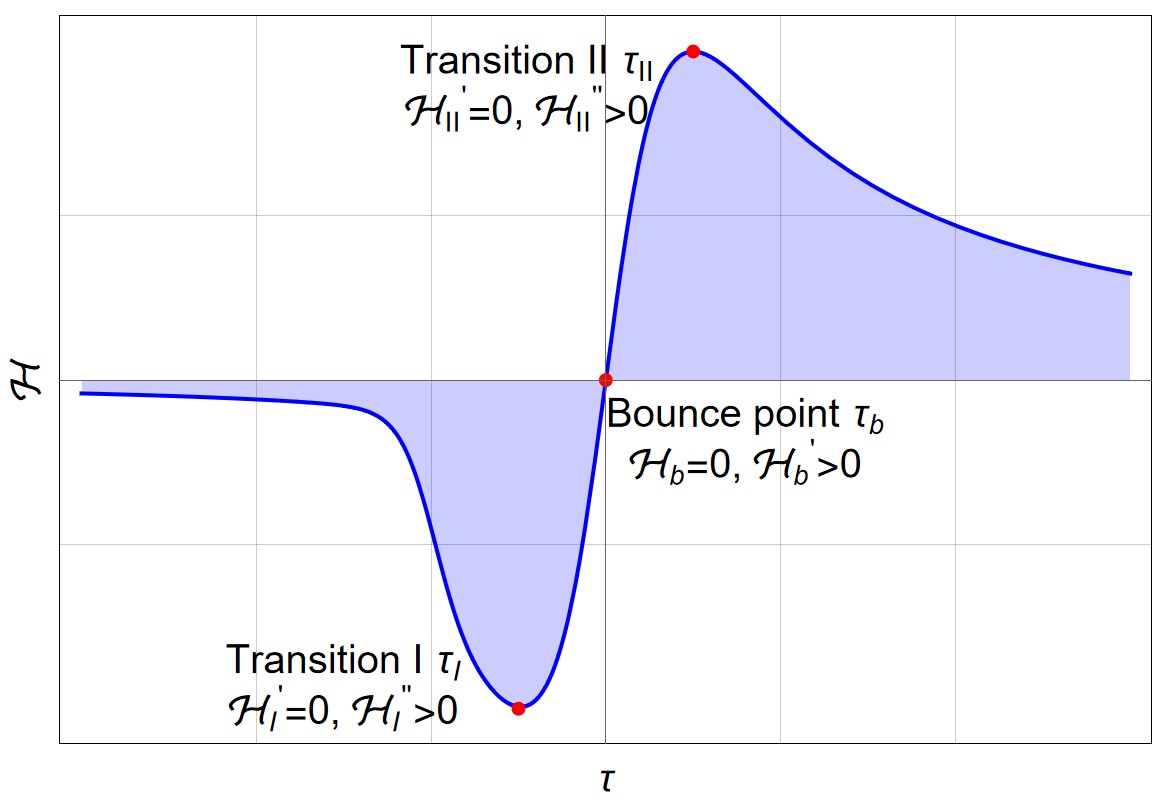}
    \caption{Illustration of the bouncing scenarios.}
    \label{fig:bounce}
\end{figure}

To analyze the structure of effective potential in bounce cosmologies, we come to the function $V^{\prime} (\tau) = 2 \mathcal{H} \mathcal{H}^{\prime} + \mathcal{H}^{\prime \prime}$. The number of peaks in the effective potential equals to the number of zero-points in $V^{\prime} (\tau)$. In non-singular bounce cosmologies, there must be a bouncing phase where the universe transits from contraction ($\mathcal{H}<0$) to expansion ($\mathcal{H}>0$). In the bouncing phase, there must be two transition points satisfying $\mathcal{H}^{\prime} = 0$, which we label as $\tau_I$ and $\tau_{II}$, as shown in Fig. \ref{fig:bounce}. It's straightforward to see from Fig. \ref{fig:bounce} that  $\mathcal{H}^{\prime \prime}(\tau_I) > 0$ and $\mathcal{H}^{\prime \prime}(\tau_{II}) < 0$, so
\begin{equation}
    V^{\prime}(\tau_I) > 0 ~,~ V^{\prime}(\tau_{II}) < 0 ~.
\end{equation}
Namely, there is a zero-point of $V^{\prime}(\tau)$ in the regime $\tau_I < \tau < \tau_{II}$ due to the mean value theorem. As a result, a peak of effective potential emerges at that place.

The emergence of the second peak is a direct relic of the dynamics governing the contracting phase. According to the previous discussion, there should be a contracting phase with $w_c > 1/3$ to bypass the anisotropic problem. In this phase
\begin{equation}
    V^{\prime}(\tau) = \frac{4(1-3w_c)}{(1+3w_c)^2 (-\tau)^3} < 0 ~.
\end{equation}
This contracting phase takes place before the first transition event $\tau = \tau_I$, where $V^{\prime}(\tau_I) > 0$. Therefore, there must be a second peak in the regime $\tau < \tau_I$ due to the mean value theorem.

We conclude that the effective potential always has at least two peaks in non-singular bounce models that is free from anisotropic stress problem. This double-peak structure of effective potential in those models leads to unique oscillatory features in energy density spectrum of GWs on high-frequency regime absent in generical inflation models. Consequently, this oscillatory signature provides a key observational discriminant between early universe scenarios.  

\section{Features of Gravitational Waves in Bouncing Cosmologies}

The dynamics of a bouncing universe introduce three pivot scales that shape the GWs spectrum. The infrared and intermediate scales, $k_{\rm IR}$ and $k_{\rm IM}$ , are those who crosses the Hubble horizons at the end of the reheating and contracting phase. The ultraviolet (UV) scale, $k_{\rm UV} \equiv \sqrt{\max |V(\tau)|}$, corresponds to the smallest perturbations, which cross the Hubble horizon exactly once. Of specific phenomenological interest is the regime $k_{\rm IR} < k_{\rm IM}$, wherein the reheating epoch contributes meaningfully to the primordial gravitational wave background. This is the scenario we shall consider throughout the paper. The spectrum of GWs exhibits generical properties on different scales:
\begin{itemize}
    \item Primordial fluctuations with $k > k_{\rm IM}$ are sub-horizon during the contracting phase. They can hardly feel the contraction of the universe, so $|\beta_k|$ is relevant to the peak-structure of effective potential only. The oscillatory feature takes place when $k \geq k_{\rm UV}$, where the Born approximation holds and \eqref{eq:betakUV} tells
    \begin{equation}
        \Omega_{\rm GW} \propto k^4 \left[ 1 + \kappa^2 \sin (\omega k/k_{\rm UV}) \right] e^{-\mu (k/k_{\rm UV} )} ~.
    \end{equation}
    Notably, the pole structure of effective potential gives an additional $k^2$ factor when performing contour integrations, so $\Omega_{\rm GW}$ scales as $k^4$ instead of $k^2$ \cite{Pi:2024kpw}. In the regime $k_{\rm IM} < k \ll k_{\rm UV}$, the solution of Eq. \eqref{eq:vkdynamical} is the Jost function, whose leading-order amplitude is $|\beta_k|^2 \simeq 1 - 4k^2 a_s^2 + \mathcal{O}(k^3)$ where $a_s$ reflects the structure of peaks. 
    \item Primordial fluctuations with $k < k_{\rm IM}$ cross the Hubble horizon during the contracting phase. Since these fluctuations remain super-horizon at the peak locations, their subsequent evolution is left essentially untouched by the double-peak structure. For fluctuations re-enter the Hubble horizon during the RD epoch, namely $k < k_{\rm IR}$, the corresponding spectra index of primoridal GWs is $n_{\rm IR} \equiv 3 - \frac{3|w_c - 1|}{1+3w_c}$ \cite{Wands:1998yp}.
    We see the spectra index satisfies $2 < n_{\rm IR} < 3$ for $w_c > 1/3$, which is in agreement with the previous analysis. On the other hand, primordial fluctuations with $k_{\rm IR} < k < k_{\rm IM}$ re-enter the horizon in reheating phase instead. The corresponding tensor spectrum is modified by a factor $|\chi_k|^2 \propto k^{2\frac{3w_{\rm rh} - 1}{3w_{\rm rh} + 1}}$ by assuming a constant effective EoS parameter $w_{\rm rh}$; see Appendix~\ref{app:reheating} for details. Therefore, the spectra index in the regime $k_{\rm IR} \lesssim k\lesssim k_{\rm IM}$ is $n_{\rm IM} \equiv n_{\rm IR} + 2 \frac{3w_{\rm rh} - 1}{3w_{\rm rh} + 1}$.
\end{itemize}

We numerically verify the asymptotic behaviors discussed above in Appendix~\ref{app:fig1-plotting}. At low-frequency regime, the spectrum exhibits a broken-power-law behavior, with spectra index $n_{\rm IR}$ on $f \leq f_{\rm IR}$ and $n_{\rm IM}$ on $f_{\rm IR} < f \leq f_{\rm IM}$. At high-frequency regimes, the spectrum slowly decreases on $f_{\rm IM} \leq f \leq f_{\rm UV}$, and then decays as $k^4e^{-\mu k}$ with a characteristic osclllations on $f \geq f_{\rm UV}$.

In the end, the amplitude of GWs spectrum can be estimated at the intermediate scale $f = f_{\rm IM}$ in Appendix~\ref{app:amplitude}
    \begin{align}
    \label{eq:amplitude}
    (\Omega_{\rm GW}h^2)(f_{\rm IM}) & \nonumber \sim 3.7 \times 10^{-33} \mathcal{A}^2 \left( \frac{H_c}{H_{\rm RD}} \right)^2   \\
    & \times \left( \frac{f_{\rm IM}}{1 {\rm Hz}} \right)^4  \left( \frac{f_{\rm IR}}{f_{\rm IM}} \right)^{\frac{1 + 3w_{\rm rh}}{3(1+w_{\rm rh})}}  ~.
    \end{align}
where $H_c$, $H_{\rm RD}$ the Hubble parameter at the end of contraction/reheating phase, respectively, $\mathcal{A}$ represents the amplitude factor of GWs from bouncing phase due to the tachyonic amplifications. 

Notably, the amplitude \eqref{eq:amplitude} can be large enough and detectable by future observations with a physically motivated model parameters for two reasons. First, the blue-tilted GWs spectrum from the contraction phase with $n_{\rm IR}$ can lead to significant growth of energy density spectrum of GWs on high-frequency regime. For instance, even a vanishingly small tensor-to-scalar ratio at Cosmic Microwave Background (CMB) regime can lead to a sizable primordial GWs signatures testable by Pulsar Timing Array (PTA) experiments at Nano-Hertz regime \cite{Vagnozzi:2020gtf,Benetti:2021uea,Vagnozzi:2023lwo}. This scale dependence has been translated into $H_c/H_{\rm RD}$ in \eqref{eq:amplitude}. A possible huge hierarchy $H_c/H_{\rm RD}$ can arise from the energy scale difference between the Null-Energy-Condition violation physics and the reheating phase, or equivalently a relatively long reheating phase, see Appendix~\ref{app:amplitude} for more details. Second, the bouncing phase may further amplify the primordial GWs due to tachyonic instabilities \cite{Cai:2011zx,Cai:2012va}.

\begin{figure*}[ht]
    \centering
    \includegraphics[width=3.2in]{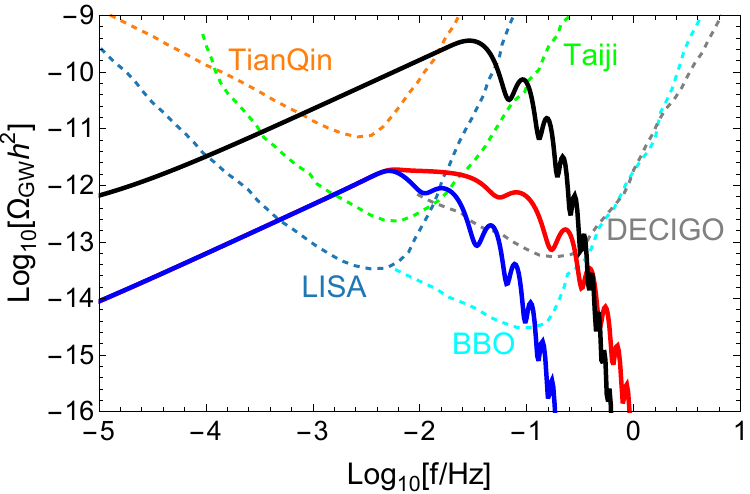}
    \includegraphics[width=3.2in]{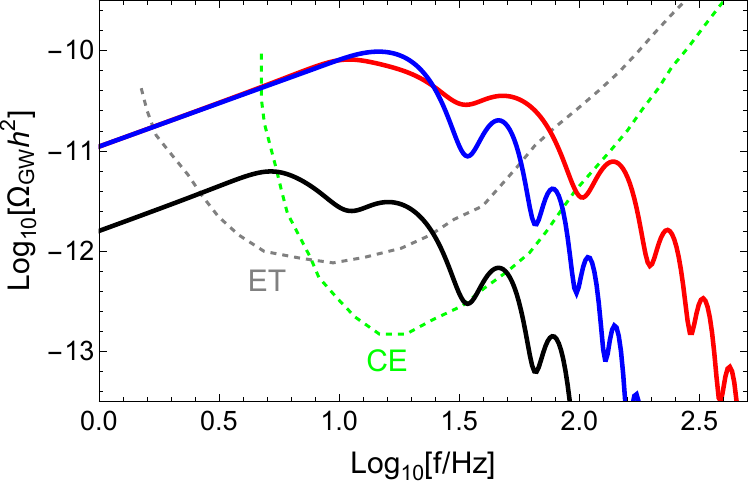}
    \caption{GWs signals (solid lines) versus experimental sensitivity curves as a function of $f$ (dashed lines). Left panel: GWs spectrum in the frequency band $10^{-5}$-$1$ Hz, with sensitivity curve of Taiji (green), TianQin (orange), LISA (light blue), BBO (cyan) and DECIGO (gray). Right panel: GWs spectrum in the frequency band $1$-$10^3$ Hz, with sensitivity curve of CE (green) and ET (gray). The GWs signals are evaluated with $w_c = 1.2$, $w_{\rm rh} = 0$ so the spectra index is $n_{\rm IM} = 0.87$ at intermediate regimes $f \leq f_{\rm IM}$. This is different from the scalar-induced GWs in inflation \cite{Domenech:2021ztg} where the spectrum has a log-dependence slope $n_{\rm GW}(f) = 3 - 2 \ln(f/f_c) $ in the IR regime $f \ll f_c$ \cite{Yuan:2019wwo}. Other values of model parameters can be found in Appendix~\ref{app:fig3-details}.}
    \label{fig:forecast}
\end{figure*}

Armed with the amplitude at $f = f_{\rm IM}$ and the scale-dependent behavior discussed above, we can translate the model parameters (summarized in Table. \ref{tab:para}) into a predicted spectrum for primordial GWs.

\begin{table*}[t]
\centering
\small
\begin{tabular}{|c|l|}
\hline
\multicolumn{1}{|l|}{Parameter} & \multicolumn{1}{c|}{Interpretations}           \\ 
\hline
$\kappa^2$   & Amplitude of oscillations; determined by amplitude of peaks \\ 
\hline
$\mu$ & Decay rate; determined by width of peaks \\ 
\hline
$\omega$ & Oscillation frequency; determined by distance of peaks \\ 
\hline
$w_c$/$w_{\rm rh}$   & EoS parameters in contraction/reheating; determine $n_{\rm IR}$ and $n_{\rm IM}$ \\ 
\hline
$H_{\rm RD}$   & Hubble parameter at the end of reheating \\ 
\hline
$H_c$ & Hubble parameter at the end of contraction \\ 
\hline
$\mathcal{A}$ & Tachyonic amplification factor \\ 
\hline
$f_{\rm IR}/f_{\rm IM}/f_{\rm UV}$ & Pivot scales defined in the main text \\ 
\hline
\end{tabular}
\caption{Parameters of primordial gravitational waves used in our analysis.}
\label{tab:para}
\end{table*}

\section{Forecast for High-frequency Gravitational-wave Experiments}

We present the resulting GWs spectrum and compare it with the experimental sensitivity curves in Fig. \ref{fig:forecast}. It is worth noting that the infrared tail of the spectrum lies beyond measurable reach. This is unsurprising given its origin: the IR pivot frequency $f_{\rm IR}$ is governed by the Hubble parameter during radiation domination via $\left( \frac{f_{\rm IR}}{1 {\rm Hz}} \right)^2 = 1.1 \times 10^{-27} \left( \frac{H_{\rm RD}}{1 {\rm GeV}} \right)$ in Appendix~\ref{app:amplitude}. The condition $H_{\rm RD} \ll M_p$ then forces $f_{\rm IR} \ll 10^{-5} {\rm Hz}$, relegating it to frequencies far lower than those we can probe.

On the other hand, high-frequency primordial GWs $ f >f_{\rm IR}$ are experimentally testable. The blue-tilted GWs spectrum on intermediate scales $f_{\rm IR} \ll f < f_{\rm IM}$ falls within the sensitivity windows of TianQin \cite{TianQin:2020hid}, Taiji \cite{Ruan:2018tsw}, and LISA \cite{LISA:2017pwj}, as illustrated in the left panel of Fig. \ref{fig:forecast}. Meanwhile, the oscillatory features emerging at higher frequencies $f \gtrsim f_{\rm UV}$ are accessible to next-generation detectors such as BBO \cite{Crowder:2005nr} and DECIGO \cite{Kawamura:2011zz}. A combined observation of the primordial GWs spectrum spanning the LISA and BBO frequency bands would therefore constitute a distinctive fingerprint of bounce cosmologies, as the specific pattern of signals across these different frequency regimes is both unique and characteristic.

Remarkably, the distinctive signal persists into higher frequency regimes. The right panel of Fig. \ref{fig:forecast} demonstrates that around $10^2$ Hz, the GWs spectrum enters the detection windows of upcoming terrestrial facilities, including Cosmic Explorer (CE) \cite{Evans:2023euw} and the Einstein Telescope (ET) \cite{ET:2025xjr}. A detection in this band, revealing the characteristic blue spectrum followed by oscillatory patterns, would offer independent and corroborating evidence for the bounce scenario.

As suggested by Eq. \eqref{eq:amplitude}, the amplitude of the gravitational wave spectrum grows with $f_{\rm IM}$. Consequently, the signal becomes prominent for large $f_{\rm IM}$, highlighting the high-frequency gravitational waves (HFGWs) as a promising observational target. While direct detection above MHz frequencies remains technically challenging and largely unexplored \cite{Aggarwal:2020olq,Aggarwal:2025noe}, recent years have witnessed a growing interest in this field, motivated by its unique significance to early universe cosmology and new physics beyond the Standard Model. Considerable attention has been devoted to not only the HFGWs detection strategies such as the Gertsenshtein effect \cite{Rubtsov:2014uga}, the electromagnetic phenomenon \cite{Li:2023tzw,Li:2025ybm}, and the atomic quantum sensors \cite{Ye:2008rva,Cai:2025fpe}, but also experimental initiatives including BAW \cite{Goryachev:2014yra}, FLASH \cite{Alesini:2019nzq,Alesini:2023qed}, ADMX \cite{ADMX:2018gho,ADMX:2019uok,ADMX:2021nhd}, and ALPHA \cite{Lawson:2019brd,ALPHA:2022rxj}. Those proposals may soon overcome the Big Bang Nucleosynthesis bound on stochastic backgrounds, offering a compelling opportunity to test our results.

\section{Conclusions}

In this Letter, we report a distinctive signature of primordial GWs in nonsingular bounce cosmologies. For bounce models that are free from the anisotropic stress problem, the effective potential inevitably develops a double-peak structure, resulting in the oscillatory patterns of GWs spectrum at high frequency regime due to the interference of GWs between the two peaks. This oscillatory pattern is unique and can serve as a distinctive signature of bounce cosmologies. Our results reveal that the energy spectrum of primordial GWs can be at reach for current and forthcoming terrestrial and space-based GWs experiments. Thus, our finding offers an appealing approach to experimentally test bounce scenarios.

Notably, our main conclusion of Eq.~\eqref{eq:amplitude} show explicitly that the amplitude of primordial GWs at high frequency after the nonsingular bounce is related to $H_{\rm RD}$, which is identified as the Hubble parameter at the end of the reheating phase. The detection of GWs signature would therefore enable an independent constraint of $H_{\rm RD}$, complementing other cosmological probes. In this regard, our findings thereby build up a promisingly novel connection between the GW astronomy and new physics potentially operating during reheating. 

Finally, the bouncing phase that connects the contracting and expanding phase may lead to richer structures than the simple two-peak scenario considered in the current Letter. Namely, multiple bouncing phases \cite{Alesci:2016xqa, Brandenberger:2017pjz}, cyclic cosmologies \cite{Steinhardt:2001st,Steinhardt:2002ih,Khoury:2003rt} and time crystal cosmologies \cite{Bains:2015gpv,Easson:2018qgr} can yields multiple peaks in the effective potential for the propagation of primordial GWs. In such cases, oscillatory signatures still persist, though their detailed forms shall depend on the number and structure of the peaks, analogous to the resonant tunneling effect. A systematic investigation of the GWs spectrum in these scenarios therefore presents a compelling direction for future work.

\begin{acknowledgments}
We are grateful to Yong Cai, Pavel Petrov, Shi Pi, Misao Sasaki, Ao Wang, Yi Wang and Shengfeng Yan for valuable comments. This work was supported in part by the National Key R\&D Program of China (2021YFC2203100). MZ is supported by NSFC (Grant No. 12503005), Sichuan Science and Technology Program (Grant No. 2026NSFSC0804), and the Fundamental Research Funds for the Central Universities Grant No. YJ202551. YFC is supported in part by NSFC (12433002), by CAS young interdisciplinary innovation team (JCTD-2022-20), by 111 Project (B23042), by CSC Innovation Talent Funds, by USTC Fellowship for International Cooperation, and by USTC Research Funds of the Double First-Class Initiative.
\end{acknowledgments}

\begin{contribution}
Y.F.C. and M.Z. designed the main structure of the project. M.Z. contributed the data analysis and generated all figures under the supervision of Y.F.C. Both authors contributed equally in writing the manuscript as well as all calculations.
\end{contribution}

\appendix
\restartappendixnumbering
\renewcommand{\theHequation}{\thesection.\arabic{equation}}
\renewcommand{\theHfigure}{\thesection.\arabic{figure}}
\renewcommand{\theHtable}{\thesection.\arabic{table}}

\section{Plotting of Figure 1 in the main text}
\label{app:fig1-plotting}

In this section, we provide details of the models that we consider in the main text.

The bouncing scenarios we consider is the Ekpyrotic bounce, in which the contraction phase is governed by a stiff matter with effective equation-of-state $w_c = -1 + 2/(3q) > 1$ \cite{Erickson:2003zm}. As a result, the conformal Hubble parameter $\mathcal{H}$ scales as
\begin{equation}
\label{eq:Hekp}
    \mathcal{H} = \frac{q}{(1-q)\tau} < 0 ~;~ \tau < 0 ~,~ 0< q <1/3 ~.
\end{equation}
In the end, the universe enters into the standard radiation-dominated epoch, where the conformal Hubble parameter scales as 
\begin{equation}
\label{eq:Hrd}
    \mathcal{H} = \frac{1}{\tau} ~,~ \tau > 0 ~.
\end{equation}
We shall also take into account the possible reheating phase. Conventionally, one defines an effective EoS parameter $w_{\rm rh}$ during the period of reheating, evaluated by the average of the instantaneous EoS parameter \cite{Martin:2010kz}. When drawing the figures, we adpot the minimal setup with $w_{\rm re} = 0$, a choice wildly applied in inflationary cosmology which can be realized by a scalar field with canonical kinetic term and a mass term around its local minimum. Therefore, the conformal Hubble parameter scales as $\mathcal{H} \simeq 2/\tau$ in reheating phase. Notice that we select $\tau = 0$ to be the bouncing point $\mathcal{H} = 0$, so the reheating phase takes place when $\tau > 0$ and $\mathcal{H}_{\rm rh} > 0$ accordingly. 

In light of this fact, we adopt the following parametrization 
\begin{align}
\label{eq:calHpara}
    \mathcal{H} \nonumber = \frac{1}{2} \frac{\tau}{\tau^2 + \tau_0^2} & \left\{ \frac{q}{1-q} \left( 1 + \tanh [\omega_c (\tau_c - \tau)] \right) \right. \\
    & \nonumber \left. +  2\left( \tanh [\omega_r (\tau_r - \tau)] - \tanh [\omega_c (\tau_c - \tau)] \right) \right. \\
    & \left. + \left( \tanh [\omega_r (\tau - \tau_r)] + 1 \right) \right\} ~;~ \tau_r \geq \tau_c ~,~ \tau_c < 0 ~,
\end{align}
where the cosmic evolution reduces to \eqref{eq:Hekp} when $\tau \ll \tau_c$ and to \eqref{eq:Hrd} when $\tau \gg \tau_r$. In addition, a straightforward Taylor expansion around $\tau = 0$ yields $\mathcal{H} \simeq 2\tau/\tau_0^2$, so this parametrization \eqref{eq:calHpara} also describe a sharp bouncing phase with $\mathcal{H}^{\prime} \simeq 2\tau_0^{-2}$. When $\tau_c = \tau_r$, the solution \eqref{eq:calHpara} depicts a scenario without reheating phase, and the value $\tau_r - \tau_c$ can indicate the duration of reheating phase.

\begin{figure}[h]
    \centering
    \includegraphics[width=0.4\linewidth]{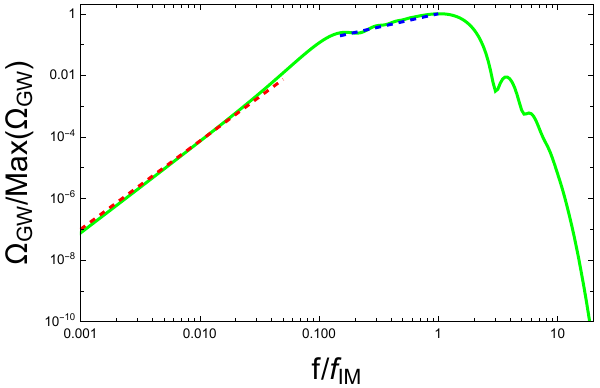}
    \includegraphics[width=0.4\linewidth]{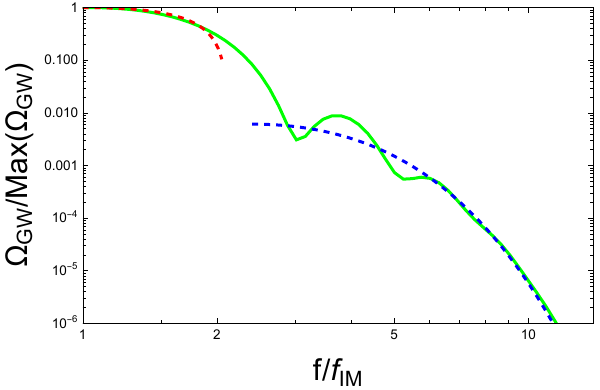}
    \caption{Left: The energy spectrum $\Omega_{\rm GW}$ (Green solid line) versus power-law function $f^{n_{\rm IR}}$ (Red dashed line) and $f^{n_{\rm UM}}$ (Blue dashed line). Right: $\Omega_{\rm GW}$ on high frequencies (Green solid line) versus the approximation $1 - 4k^2 a_s^2$ and the function $f^4e^{-\mu f}$ (Blue dashed line).}
    \label{fig:PGWtemplete}
\end{figure}

We use the following parameter sets to draw Fig. 1 in the main text:
\begin{equation}
    w_c = 1.2 ~,~ w_{\rm rh} = 0 ~;~ \omega_c = 5 ~,~ \omega_r = 0.5 ~;~ \tau_0 = 1/2 ~,~ \tau_c = -1 ~,~ \tau_r = 8 ~.
\end{equation}
The same parameters are used to numerically evaluate $\Omega_{\rm GW}$, and we show the result in Fig. \ref{fig:PGWtemplete}. The asymptotic behaviors in Fig. \ref{fig:PGWtemplete} are plotted via 
\begin{equation}
    \frac{\Omega_{\rm GW}}{\max |\Omega_{\rm GW}|} = 40 \left( \frac{f}{f_{\rm IM}} \right)^{\frac{66}{23}} ~,~  \text{ Red  dashed  line in left channel} ~,
\end{equation}
\begin{equation}
    \frac{\Omega_{\rm GW}}{\max |\Omega_{\rm GW}|} =  \left( \frac{f}{f_{\rm IM}} \right)^{\frac{20}{23}} ~,~  \text{ Blue dashed  line in left channel} ~,
\end{equation}
\begin{equation}
    \frac{\Omega_{\rm GW}}{\max |\Omega_{\rm GW}|} = 1-0.9^2 \left( \frac{f}{f_{\rm IM}} -1 \right)^{2} ~,~  \text{ Red  dashed  line in  right channel} ~,
\end{equation}
\begin{equation}
    \frac{\Omega_{\rm GW}}{\max |\Omega_{\rm GW}|} = 0.1^2 \left( \frac{f}{f_{\rm IM}} \right)^{4} e^{-\frac{5}{3} \frac{f}{f_{\rm IM}}} ~,~  \text{ Blue  dashed  line in  right channel} ~.
\end{equation}
We see that the numerical result indeed satisfies the asymptotic behaviors discussed in the main text. 

Finally, the reference scales in Fig. \ref{fig:PGWtemplete} are chosen as $\mathcal{H}_{\ast} = |\mathcal{H}|_{\rm max}$ and $\tau_{\ast} = \mathcal{H}_{\ast}^{-1}$.

\section{Physical origin of the oscillations in GWs energy density spectrum}
\label{app:oscillations}

In this section, we explain the physical mechanism producing the oscillations in the GWs spectrum. The propagation of primordial tensor fluctuations is governed by
\begin{align}
\label{eq:app_vkdynamical}
    v_k^{\prime \prime} + \left( k^2 - V(\tau) \right) v_k = 0 ~,
\end{align}
where 
\begin{equation}
    V(\tau) \equiv a^{\prime \prime}/a ~,
\end{equation}
is the effective potential. On high-frequency regime
\begin{equation}
    k^2 \gg |V(\tau)| ~,
\end{equation}
the equation \eqref{eq:app_vkdynamical} can be resolved perturbatively as a series of $|V|/k^2$. We write \eqref{eq:app_vkdynamical} as
\begin{equation}
v_k''+k^2v_k=V(\tau)v_k ~.
\end{equation}
When the right-hand side is treated perturbatively, the solution is
\begin{equation}
v_k(\tau) = \frac{e^{-ik\tau}}{\sqrt{2k}} + \int_{-\infty}^{\tau}d\tau'\, G_k(\tau,\tau')V(\tau')v_k(\tau') ~,
\end{equation}
with the retarded Green's function of the free operator,
\begin{equation}
G_k(\tau,\tau') = \frac{\sin[k(\tau-\tau')]}{k} \Theta(\tau-\tau') ~.
\end{equation}
The leading-order result is acquired by taking $v_k(\tau')$ inside the integral as the plane wave solution $e^{-ik\tau'}/\sqrt{2k}$, known as the Born approximation. In the late-time limit, we obtain
\begin{equation}
\lim_{\tau \to +\infty}v_k \simeq \frac{1}{\sqrt{2k}} \left[ \left(1 + \alpha_k\right) e^{-ik\tau} + \beta_k e^{ik\tau} \right],
\end{equation}
with
\begin{equation}
\beta_k \simeq -\frac{i}{2k} \int_{-\infty}^{+\infty}d\tau\, V(\tau)e^{-2ik\tau} ~.
\label{eq:Born_beta}
\end{equation}
Namely, in the high-frequency regime, the production of relic gravitons is controlled by the Fourier component of the effective potential at frequency $2k$.

Notably, the Bogoliubov coefficient $\beta_k$ corresponds to the particle production of gravitons, so the energy density spectrum of produced gravitons are 
\begin{equation}
    \Omega_{\rm GW} = \frac{k^4 |\beta_k|^2}{3\pi^2M_p^2H^2a^4} ~,
\end{equation}
with $\rho_c \equiv 3H_0^2 M_p^2$ the critical energy density and $H_0$ the Hubble parameter today. Therefore, the interference pattern in the Bogoliubov coefficient $\beta_k$ leads to the oscillations in the GW energy density spectrum, with a frequency inversely proportional to the separation of the two-peaks in the effective potential $V(\tau)$.

For illustrative purpose, we parametrize the double-peak in the effective potential as two localized Lotentzian fucntions:
\begin{equation}
\label{eq:app_VLorentz}
    V(\tau) = A_2^2 L(\tau; \tau_2,\Delta \tau) - A_1^2 L(\tau;\tau_1,\Delta \tau) ~,~ L(\tau; \tau',\Delta\tau)=\frac{1}{\pi}\frac{\Delta\tau}{(\tau-\tau')^2+\Delta\tau^2} ~,
\end{equation}
where $A_1$, $A_2$, $\tau_1$, $\tau_2$ label the amplitude and location of the two peaks. The resulting Bogoliubov coefficient is
\begin{equation}
    |\beta_k|^2 \simeq \frac{e^{-4k\Delta\tau}}{4k^2}\left[ A_2^4+A_1^4-2A_1^2A_2^2 \cos\left(2k(\tau_2-\tau_1)\right) \right] ~.
\end{equation}
Thus, the oscillatory frequency in the GWs spectrum is $f_{\rm osci} = (2a_0 |\tau_2 - \tau_1|)^{-1}$, inversely proportional to the distance of the two peaks, as expected.

We conclude that the double-peak structure of effective potential in non-singular bouncing cosmology leads to unique oscillatory features in energy density spectrum of GWs with a frequency inversely proportional to the distance of the two peaks on high-frequency regime. 

In the end, we comment on the location of the oscillations. The analyze relies on the high-frequency assumption $k^2 \gg |V(\tau)|$. In the main text, we have defined the characteristic scale $k_{\rm UV} \equiv \sqrt{\max |V(\tau)|}$, so the oscillatory feature takes place when $k \geq k_{\rm UV}$. In terms of frequency ,the oscillations in the GWs spectrum start at $f \simeq f_{\rm UV} = k_{\rm UV}/(2\pi a_0)$.

\section{Evolutions of primordial GWs in reheating phase}
\label{app:reheating}

In this section we explain the behavior of primordial GWs in reheating phase. The readers are referred to Ref. \cite{DEramo:2019tit,Haque:2021dha} for more technical details. The modofication of reheating phase is encoded in the tensor transfer function $\chi_k(\tau)$, defined as
\begin{equation}
    h_k(\tau) = h_k^{\rm RH} \chi_k(\tau) ~,
\end{equation}
with $h_k^{\rm RH}$ the tensor fluctuation at the beginning of reheating. This quantity obeys the same differential equation as $h_k$:
\begin{equation}
    \chi_k^{\prime \prime} + 2 \frac{a^{\prime}}{a} \chi_k^{\prime} + k^2 \chi_k = 0 ~.
\end{equation}
Assuming the reheating phase is governed by an effective equation-of-state (EoS) parameter $w_{\rm rh}$, then the Hubble parameter scales as $H \propto a^{-3(1+ w_{\rm rh})/2}$, and the general solution of transfer function is the Bessel function
\begin{equation}
\label{eq:chik}
    \chi_k \propto a^{-\frac{3}{4}(1-w_{\rm rh})} \left[ C_k J_{\nu_{\rm rh}} \left( \frac{k}{k_{\rm IR}} a^{\gamma} \right) + D_k J_{-\nu_{\rm rh}} \left( \frac{k}{\gamma k_{\rm IR}} a^{\gamma} \right) \right] ~,
\end{equation}
\begin{equation}
    \nu_{\rm rh} \equiv \frac{3(1-w_{\rm rh})}{2(1+3w_{\rm rh})} ~,~ \gamma \equiv \frac{1+3w_{\rm rh}}{2} ~.
\end{equation}
This solution holds only if $-1/3 < w_{\rm rh} < 1$, a quite generic condition as indicated by \cite{Martin:2010kz}. The parameter $k_{\rm IR}$ represents the scale re-enter the horizon at the end of reheating. Notice that, small-scale primordial fluctuations that never cross the horizon shall be correctly regularized to eliminate the vacuum contributions. We introduce another characteristic scale $k > k_{\rm UV}$ to denote the regime where the effect of regularization becomes important. So, \eqref{eq:chik} is valid only for $k < k_{\rm UV}$, and primordial fluctuations on those scales receive negligible contributions from the subsequent RD epoch, as they re-enter the horizon during reheating. As a result, the primordial energy spectrum receives an correction factor $\left| \chi_k \right|^2$ on $k < k_{\rm UV}$.

In IR regime $k \ll k_{\rm IR}$, the modifications shall be negligible as those fluctuations become sub-horizon before reheating. This condition fixes the coefficients in \eqref{eq:chik} via
\begin{equation}
    \lim_{k/k_{\rm IR} \to 0} \chi_k = 1 ~.
\end{equation}
resulting in 
\begin{equation}
    \chi_k \simeq \Gamma(\nu_{\rm rh} + 1) \left( \frac{k}{2\gamma k_{\rm IR}} \right)^{-\nu_{\rm rh}} J_{\nu_{\rm rh}} \left( \frac{k}{\gamma k_{\rm IR}} \right) ~,
\end{equation}
at today where we set the $a_0 \equiv a_{\rm today} = 1$.Notably, in the intermediate regime $k_{\rm IR} < k < k_{\rm UV}$, the Bessel function asymptotics to
\begin{equation}
\label{eq:Jasym}
    |J_{\alpha}(z)| \simeq \sqrt{\frac{2}{\pi z}} \cos \left( z - (2\alpha + 1) \frac{\pi}{4} \right) ~,
\end{equation}
so the mean value of spectrum develops an approximate power-law behavior
\begin{equation}
\label{eq:nTIM}
    n_{T,IM} = n_{T,{\rm IR}} + 2 \frac{3w_{\rm rh} - 1}{3w_{\rm rh} + 1} ~.
\end{equation}
Notably, the oscillating term shall contribute a constant value after we averaging the Bessel function over time.

\section{Amplitude of GW spectrum}
\label{app:amplitude}
In this section we briefly explain how we estimate the amplitude of $\Omega_{\rm GW}$. We label the end of contraction phase as $\tau_{\rm c}$. Notice that $\tau_{\rm c}$ labels the point where the constant $w_c$ assumption is about to break, and one shall not confuse it with $\tau_I$, the first transition point. Also, we are interested in the case $k_{\rm IM} > k_{\rm IR}$, otherwise, the reheating phase doesn't introduce interesting observational signatures.

We can analytically track $P_{T}(k_{\rm IM})$ at $\tau = \tau_{\rm IM}$ via
\cite{Wands:1998yp}
\begin{equation}
    P_{T}(k_{\rm IM},\tau_{c}) = \frac{2^{2|\nu|-1} \Gamma^2(|\nu|)}{(2\nu-1)^2\Gamma^2(3/2)} \left( \frac{H(\tau_{c})}{2\pi M_p^2} \right)^2 (-k_{\rm IM}\tau_{\rm c})^{3-2|\nu|} ~,~ \nu \equiv \frac{3(w_c - 1)}{2(1+3w_c)} ~.
\end{equation}

Since we asssumed $k_{\rm IM} > k_{\rm IR}$, $k_{\rm IM}$ re-enters the horizon at reheating phase at a time $\tau_{\rm IM}$, and the corresponding tensor fluctuations receive negligible modifications in the bouncing phase, then we can estimate
\begin{equation}
    P_{T,{\rm today}}(k_{\rm IM}) = P_{T}(k_{\rm IM},\tau_{\rm IM}) \left( \frac{a_0}{a(\tau_{\rm IM})} \right)^{-2} \simeq P_{T}(k_{\rm IM},\tau_c) \left( \frac{a_0}{a(\tau_{\rm IM})} \right)^{-2} ~.
\end{equation}
In addition, the energy spectrum at today is related to $P_T$ as
\begin{equation}
\label{eq:OGWtoday}
    \Omega_{\rm GW}(k_{\rm IM}) \simeq \frac{1}{12} \left( \frac{k_{\rm IM}}{a_0 H_0} \right)^2 P_{T,{\rm today}}(k_{\rm IM}) ~,
\end{equation}
where $a_0$, $H_0$ the scale factor and Hubble parameter today, so
\begin{equation}
    \Omega_{\rm GW}(k_{\rm IM}) \simeq \frac{1}{12}  \frac{(k_{\rm IM}/a_0)^4}{H_0^2M_p^2}  \frac{2^{2|\nu|-3} \Gamma^2(|\nu|)}{(2\nu-1)^2\Gamma^2(3/2) \pi^2} \left( \frac{H(\tau_c)}{H(\tau_{\rm IM})} \right)^2 \left( \frac{2}{1+3w_c} \right)^{3-2|\nu|} ~,
\end{equation}
where we've used the horizon crossing condition $-k_{\rm IM}\tau_c = 2/(1+3w_c)$. 

We can simplify $H(\tau_c)/H(\tau_{\rm IM})$ by noticing that
\begin{equation}
    \frac{H(\tau_c)}{H(\tau_{\rm IM})} = \frac{H(\tau_c)}{H(\tau_{\rm IR})} \frac{H(\tau_{\rm IR})}{H(\tau_{\rm IM})} = \frac{H(\tau_c)}{H(\tau_{\rm IR})} \left( \frac{k_{\rm IR}}{k_{\rm IM}} \right)^{\frac{1 + 3w_{\rm rh}}{3(1+w_{\rm rh})}} ~.
\end{equation}
Notice that, $|H(\tau_c)|$ also labels the highest energy scale in contraction phase, and $H(\tau_{\rm IR})$ is the Hubble parameter at the beginning of RD epoch. Let us introduce $|H(\tau_c)|$ as $H_c$ and  $H(\tau_{\rm IR})$ as $H_{\rm RD}$, then the $|H_c|/H_{\rm RD}$ characterizes the ratio of energy scales between contraction and RD epochs. 

We estimate $2^{3-2|\nu|}/(1-2\nu)^{1-2\nu}$ as 3, since it ranges from 1 to 8.3 when $|\nu| \leq 1/2$. We also notice that $\Gamma(|\nu|)$ and $2/(1+3w_c)$ is usually order unity, unless we assign some extreme value $|\nu| \ll 1$ or $w_c \gg 1$. This leads to
\begin{equation}
    \label{eq:OGWsimple}
    (\Omega_{\rm GW}h^2)(f_{\rm IM}) \sim 3.7 \times 10^{-33} \left( \frac{H_c}{H_{\rm RD}} \right)^2 \left( \frac{f_{\rm IM}}{1 {\rm Hz}} \right)^4 \left( \frac{f_{\rm IR}}{f_{\rm IM}} \right)^{\frac{1 + 3w_{\rm rh}}{3(1+w_{\rm rh})}} ~.
\end{equation}

Unfortunately, the estimation may ignore an important contributions from bouncing phase. Especially, primordial fluctuations may experience an exponentially super-horizon growth due to tachyonic instabilities during bouncing phase \cite{Cai:2012va}. In this case, \eqref{eq:OGWsimple} shall acquire an large amplification factor. For example, the amplification factor for scalar is designed to be around $\mathcal{O}(10^2)$ larger than tensor to suppress the tensor-to-scalar ratio \cite{Cai:2011zx}. We introduce a new parameter $\mathcal{A}$ to characterize the amplification of tensor fluctuations during bouncing phase, and the final result shall be
\begin{equation}
    (\Omega_{\rm GW}h^2)(f_{\rm IM}) \sim 3.7 \times 10^{-33} \mathcal{A}^2 \left( \frac{H_c}{H_{\rm RD}} \right)^2 \left( \frac{f_{\rm IM}}{1 {\rm Hz}} \right)^4 \left( \frac{f_{\rm IR}}{f_{\rm IM}} \right)^{\frac{1 + 3w_{\rm rh}}{3(1+w_{\rm rh})}} ~.
\end{equation}
Since $\mathcal{A}$ represents the influence on primordial GWs from the bouncing phase, the measurement of $\mathcal{A}$ from GW experiments can probe the microscopic physics of bouncing phase and the underlying new physics of NEC violation.

In the end, we comment that if we assume the standard $\Lambda$CDM scenario, the relativistic degree of freedom in RD epoch is $g_{\ast} = 106.75$ and $H_{\rm RD}$ is related to $f_{\rm IR}$ via
\begin{equation}
    \left( \frac{f_{\rm IR}}{1 {\rm Hz}} \right)^2 = 1.1 \times 10^{-27} \left( \frac{H_{\rm RD}}{1 {\rm GeV}} \right) ~.
\end{equation}

\section{A generic parametrization of primordial GWs and plotting details of Fig. 3}
\label{app:fig3-details}

In light of the properties of primordial GWs discussed in the main text, we capture the essense of GWs spectrum as
\begin{align}
\label{eq:OGWtemplete}
    (\Omega_{\rm GW}h^2) & \nonumber = 3.7 \times 10^{-33} \mathcal{A}^2 \left( \frac{H_c}{H_{\rm RD}} \right)^2 \left( \frac{f_{\rm IM}}{1 {\rm Hz}} \right)^4 \left( \frac{f_{\rm IR}}{f_{\rm IM}} \right)^{\frac{1 + 3w_{\rm rh}}{3(1+w_{\rm rh})}} \\
    & \nonumber \times \left\{ \left( \frac{f}{f_{\rm IR}} \right)^{n_{\rm IR}} \left( \frac{f_{\rm IR}}{f_{\rm IM}} \right)^{n_{\rm IM}} [1 - \mathcal{D}(f;f_{\rm IR})] \right. \\
    & \nonumber \left. + \left( \frac{f}{f_{\rm IM}} \right)^{n_{\rm IM}} [ \mathcal{D}(f;f_{\rm IR}) - \mathcal{D}(f;f_{\rm IM})] \right. \\
    & \nonumber \left. + \frac{1 - 4a_s^2 \frac{f^2}{f_{\rm UV}^2}}{1 - 4a_s^2 \frac{f_{\rm IM}^2}{f_{\rm UV}^2}} [ \mathcal{D}(f;f_{\rm IM}) - \mathcal{D}(f;f_{\rm UV})] \right. \\
    &  + \left. \frac{1 - 4a_s^2}{1 - 4a_s^2 \frac{f_{\rm IM}^2}{f_{\rm UV}^2}} e^{-\mu \frac{f-f_{\rm UV}}{f_{\rm UV}} }  \mathcal{D}(f;f_{\rm UV}) \left[ 1 + \kappa^2 \sin \left(\omega \frac{f - f_{\rm UV}}{f_{\rm UV}} \right) \right] \right\} ~,
\end{align}
where
\begin{equation}
    \mathcal{D}(f;f_0) \equiv \frac{1 + \tanh [(f-f_0)/f_{\#}]}{2} ~,
\end{equation}
are introduced to smoothly connect each pieces.

We plot the GWs signal in Fig.3 using \eqref{eq:OGWtemplete}with the following parameters
\begin{equation}
    \mu = 0.5 ~,~ \kappa = 0.8 ~,~ \omega = 2 ~,
\end{equation}
\begin{equation}
    w_c = 1.2 ~,~ w_{\rm rh} = 0 ~,~ H_c = 10^{16} {\rm GeV} ~,~ a_s = 0.4 ~.
\end{equation}

The rest parameters are summarized in Tab. \ref{tab:modelparas}.

\begin{table}[h]
\begin{tabular}{|l|c|c|}
\hline
           & \multicolumn{1}{c|}{\begin{tabular}[c]{@{}c@{}}Left panel\\ $H_{\rm rd} = 10^2 {\rm GeV}$, $\mathcal{A} = 300$\end{tabular}} & \multicolumn{1}{c|}{\begin{tabular}[c]{@{}c@{}}Right panel \\ $H_{\rm rd} = 2 \times 10^5 {\rm GeV}$, $\mathcal{A} = 1$ \end{tabular}} \\
\hline
Red line   & $f_{\rm IM} = 5 \times 10^{-3} {\rm Hz}$ , $f_{\rm UV} = 5 \times 10^{-2} {\rm Hz}$    & $f_{\rm IM} = 10$ , $f_{\rm UV} = 30 {\rm Hz}$     \\ \hline
Blue line  & $f_{\rm IM} = 5 \times 10^{-3} {\rm Hz}$ , $f_{\rm UV} = 1 \times 10^{-2} {\rm Hz}$    & $f_{\rm IM} = 10$ , $f_{\rm UV} = 10 {\rm Hz}$     \\ \hline
Black line & $f_{\rm IM} = 2 \times 10^{-2} {\rm Hz}$ , $f_{\rm UV} = 2 \times 10^{-2} {\rm Hz}$    & $f_{\rm IM} = 5$ , $f_{\rm UV} = 10 {\rm Hz}$     \\ \hline
\end{tabular}
\caption{Parameters that are used to plot Fig.3 in the main text.}
\label{tab:modelparas}
\end{table}

\bibliography{ref}
\bibliographystyle{aasjournalv7}

\end{document}